\documentclass[12pt,a4paper]{article}
\usepackage{ifthen} % for conditional statements
\usepackage{subcaption}            % for multiple sub-figures in a single float
\newboolean{pdflatex}
\setboolean{pdflatex}{true} % False for eps figures 

\newboolean{articletitles}
\setboolean{articletitles}{true} % False removes titles in references

\newboolean{uprightparticles}
\setboolean{uprightparticles}{false} %True for upright particle symbols

\def\paperauthors{Lucas Arthur Foreman} % Leave as is for PAPER, CONF and FIGURE
\def\paperasciititle{Measurements of time-dependent b-hadron properties in LHCb} % Set ASCII title here !! MAKE sure it's only ASCII characters !! 
\def\papertitle{Measurements of time-dependent \bquark-hadron properties in LHCb} % Latex formatted title
\def\paperkeywords{{High Energy Physics}, {LHCb}} % Comma separated list
\def\papercopyright{\the\year\ CERN for the benefit of the LHCb collaboration} % new since 9/Apr/2018

\def\paperlicenceurl{https://creativecommons.org/licenses/by/4.0/}

\newif\ifEnableSectionTOCLinks
\EnableSectionTOCLinksfalse % deactivated

\usepackage[top=1in, bottom=1.25in, left=1in, right=1in]{geometry}

\usepackage{microtype}
\usepackage{lineno}  % for line numbering during review
\usepackage{xspace} % To avoid problems with missing or double spaces after
\usepackage{caption} %these three command get the figure and table captions automatically small

\usepackage{graphicx}  % to include figures (can also use other packages)
\usepackage{color}
\usepackage{colortbl}
\graphicspath{{./figs/}} % Make Latex search fig subdir for figures
\usepackage{amsmath} % Adds a large collection of math symbols
\usepackage{amssymb}
\usepackage{amsfonts}
\usepackage{upgreek} % Adds in support for greek letters in roman typeset

\usepackage[normalem]{ulem} % for \sout strikeout in template tables

\newcommand*\patchAmsMathEnvironmentForLineno[1]{%
\expandafter\let\csname old#1\expandafter\endcsname\csname #1\endcsname
\expandafter\let\csname oldend#1\expandafter\endcsname\csname
end#1\endcsname
 \renewenvironment{#1}%
   {\linenomath\csname old#1\endcsname}%
   {\csname oldend#1\endcsname\endlinenomath}%
}
\newcommand*\patchBothAmsMathEnvironmentsForLineno[1]{%
  \patchAmsMathEnvironmentForLineno{#1}%
  \patchAmsMathEnvironmentForLineno{#1*}%
}
\AtBeginDocument{%
\patchBothAmsMathEnvironmentsForLineno{equation}%
\patchBothAmsMathEnvironmentsForLineno{align}%
\patchBothAmsMathEnvironmentsForLineno{flalign}%
\patchBothAmsMathEnvironmentsForLineno{alignat}%
\patchBothAmsMathEnvironmentsForLineno{gather}%
\patchBothAmsMathEnvironmentsForLineno{multline}%
\patchBothAmsMathEnvironmentsForLineno{eqnarray}%
}

\usepackage[pdfauthor={\paperauthors},
            pdftitle={\paperasciititle},
            pdfkeywords={\paperkeywords}]{hyperref}
\usepackage{hyperxmp}
\hypersetup{
    pdfcopyright={Copyright (C) \papercopyright},
    pdflicenseurl={\paperlicenceurl}
}
\usepackage[colorinlistoftodos,textsize=scriptsize]{todonotes}

\usepackage[bottom,flushmargin,hang,multiple]{footmisc}

\usepackage[all]{hypcap} % Internal hyperlinks to floats.

\usepackage{xspace} 
\usepackage{upgreek}

\def\lhcb   {\mbox{LHCb}\xspace}

\def\MagUp {\mbox{\em Mag\kern -0.05em Up}\xspace}

\ifdefined\ifuprightparticles
\else
\newboolean{uprightparticles}
\setboolean{uprightparticles}{false} %True for upright particle symbols
\fi

\ifthenelse{\boolean{uprightparticles}}%
{

 \def\Pmu         {\ensuremath{\upmu}\xspace}

 \def\Ppi         {\ensuremath{\uppi}\xspace}                 
                  
 \def\Prho        {\ensuremath{\uprho}\xspace}

 \def\Ppsi        {\ensuremath{\uppsi}\xspace}

 \def\PDelta      {\ensuremath{\Delta}\xspace}                 
 \def\PXi         {\ensuremath{\Xi}\xspace}                 
 \def\PLambda     {\ensuremath{\Lambda}\xspace}                 
 \def\PSigma      {\ensuremath{\Sigma}\xspace}                 
 \def\POmega      {\ensuremath{\Omega}\xspace}                 
 \def\PUpsilon    {\ensuremath{\Upsilon}\xspace}
 \let\oldPi\Pi
 \def\PPi         {\ensuremath{\oldPi}\xspace}

 \def\PB      {\ensuremath{\mathrm{B}}\xspace}                 
 \def\PD      {\ensuremath{\mathrm{D}}\xspace}                 
 \def\PJ      {\ensuremath{\mathrm{J}}\xspace}                 
 \def\PK      {\ensuremath{\mathrm{K}}\xspace}                 
 \def\Pb      {\ensuremath{\mathrm{b}}\xspace}                 
 \def\Pc      {\ensuremath{\mathrm{c}}\xspace}                 
 \def\Pd      {\ensuremath{\mathrm{d}}\xspace}                 

 \def\Ps      {\ensuremath{\mathrm{s}}\xspace}

 \def\thebaroffset{0.0em}
}
{

 \def\Pmu         {\ensuremath{\mu}\xspace}

 \def\Ppi         {\ensuremath{\pi}\xspace}                 
                  
 \def\Prho        {\ensuremath{\rho}\xspace}

 \def\Ppsi        {\ensuremath{\psi}\xspace}                 
                  
 \mathchardef\PDelta="7101
 \mathchardef\PXi="7104
 \mathchardef\PLambda="7103
 \mathchardef\PSigma="7106
 \mathchardef\POmega="710A
 \mathchardef\PUpsilon="7107
 \mathchardef\PPi="7105
 \def\PB      {\ensuremath{B}\xspace}                 
 \def\PD      {\ensuremath{D}\xspace}                 
 \def\PJ      {\ensuremath{J}\xspace}                 
 \def\PK      {\ensuremath{K}\xspace}                 
 \def\Pb      {\ensuremath{b}\xspace}                 
 \def\Pc      {\ensuremath{c}\xspace}                 
 \def\Pd      {\ensuremath{d}\xspace}                 

 \def\Ps      {\ensuremath{s}\xspace}

 \def\thebaroffset{0.18em}
}
\newcommand{\offsetoverline}[2][\thebaroffset]{\kern #1\overline{\kern -#1 #2}}%

\makeatletter
\ifcase \@ptsize \relax% 10pt
  \newcommand{\miniscule}{\@setfontsize\miniscule{4}{5}}% \tiny: 5/6
\or% 11pt
  \newcommand{\miniscule}{\@setfontsize\miniscule{5}{6}}% \tiny: 6/7
\or% 12pt
  \newcommand{\miniscule}{\@setfontsize\miniscule{5}{6}}% \tiny: 6/7
\fi
\makeatother

\DeclareRobustCommand{\optbar}[1]{\shortstack{{\miniscule (\rule[.5ex]{1.25em}{.18mm})}
  \\ [-.7ex] $#1$}}

\def\mup        {{\ensuremath{\Pmu^+}}\xspace}
\def\mun        {{\ensuremath{\Pmu^-}}\xspace} % muon negative (\mum is taken)

\def\ellm       {{\ensuremath{\ell^-}}\xspace}
\def\ellp       {{\ensuremath{\ell^+}}\xspace}

\def\dquark    {{\ensuremath{\Pd}}\xspace}

\def\squark    {{\ensuremath{\Ps}}\xspace}

\def\cquark    {{\ensuremath{\Pc}}\xspace}
\def\cquarkbar {{\ensuremath{\overline \cquark}}\xspace}

\def\bquark    {{\ensuremath{\Pb}}\xspace}
\def\bquarkbar {{\ensuremath{\overline \bquark}}\xspace}

\def\pion   {{\ensuremath{\Ppi}}\xspace}

\def\pip    {{\ensuremath{\pion^+}}\xspace}
\def\pim    {{\ensuremath{\pion^-}}\xspace}

\def\rhomeson {{\ensuremath{\Prho}}\xspace}
\def\rhoz     {{\ensuremath{\rhomeson^0}}\xspace}

\def\kaon    {{\ensuremath{\PK}}\xspace}
\def\KorKbar {\kern \thebaroffset\optbar{\kern -\thebaroffset \PK}{}\xspace}

\def\Kp      {{\ensuremath{\kaon^+}}\xspace}

\def\KS      {{\ensuremath{\kaon^0_{\mathrm{S}}}}\xspace}

\def\Kstarz  {{\ensuremath{\kaon^{*0}}}\xspace}

\def\D       {{\ensuremath{\PD}}\xspace}

\def\DorDbar {\kern \thebaroffset\optbar{\kern -\thebaroffset \PD}\xspace}

\def\Dp      {{\ensuremath{\D^+}}\xspace}
\def\Dm      {{\ensuremath{\D^-}}\xspace}

\def\DpDm    {\ensuremath{\Dp {\kern -0.16em \Dm}}\xspace}

\def\Dsm     {{\ensuremath{\D^-_\squark}}\xspace}

\def\B       {{\ensuremath{\PB}}\xspace}

\def\BorBbar {\kern \thebaroffset\optbar{\kern -\thebaroffset \PB}\xspace}

\def\Bd      {{\ensuremath{\B^0}}\xspace}

\def\BdorBdbar {\kern \thebaroffset\optbar{\kern -\thebaroffset \Bd}\xspace}
\def\Bu      {{\ensuremath{\B^+}}\xspace}

\def\Bp      {{\ensuremath{\Bu}}\xspace}

\def\Bs      {{\ensuremath{\B^0_\squark}}\xspace}

\def\BsorBsbar {\kern \thebaroffset\optbar{\kern -\thebaroffset \Bs}\xspace}

\def\Bds     {{\ensuremath{\B_{(\squark)}^0}}\xspace}

\def\jpsi     {{\ensuremath{{\PJ\mskip -3mu/\mskip -2mu\Ppsi}}}\xspace}

\def\Y#1S{\ensuremath{\PUpsilon{(#1S)}}\xspace}

\def\LorLbar     {\kern \thebaroffset\optbar{\kern -\thebaroffset \PLambda}\xspace}

\def\to                 {\ensuremath{\rightarrow}\xspace}

\def\CP                {{\ensuremath{C\!P}}\xspace}

\newcommand{\phis}{{\ensuremath{\phi_{\squark}}}\xspace}

\newcommand{\mistag}{\ensuremath{\omega}\xspace}

\newcommand{\etag}{{\ensuremath{\varepsilon_{\mathrm{tag}}}}\xspace}

\newcommand{\effeff}{\ensuremath{\varepsilon_{\mathrm{eff}}}\xspace}

\def\AT#1     {\ensuremath{A_{\mathrm{T}}^{#1}}\xspace}           % 2

\def\C#1      {\ensuremath{\mathcal{C}_{#1}}\xspace}                       % 9
\def\Cp#1     {\ensuremath{\mathcal{C}_{#1}^{'}}\xspace}                    % 7
\def\Ceff#1   {\ensuremath{\mathcal{C}_{#1}^{\mathrm{(eff)}}}\xspace}        % 9  
\def\Cpeff#1  {\ensuremath{\mathcal{C}_{#1}^{'\mathrm{(eff)}}}\xspace}       % 7
\def\Ope#1    {\ensuremath{\mathcal{O}_{#1}}\xspace}                       % 2
\def\Opep#1   {\ensuremath{\mathcal{O}_{#1}^{'}}\xspace}                    % 7

\newcommand{\aunit}[1]{\ensuremath{\text{\,#1}}}       
\newcommand{\tev}{\aunit{Te\kern -0.1em V}\xspace}
\newcommand{\gev}{\aunit{Ge\kern -0.1em V}\xspace}
\newcommand{\mev}{\aunit{Me\kern -0.1em V}\xspace}
\newcommand{\kev}{\aunit{ke\kern -0.1em V}\xspace}
\newcommand{\ev}{\aunit{e\kern -0.1em V}\xspace}
 
\newcommand{\mevc}{\ensuremath{\aunit{Me\kern -0.1em V\!/}c}\xspace}
\newcommand{\gevc}{\ensuremath{\aunit{Ge\kern -0.1em V\!/}c}\xspace}
\newcommand{\mevcc}{\ensuremath{\aunit{Me\kern -0.1em V\!/}c^2}\xspace}
\newcommand{\gevcc}{\ensuremath{\aunit{Ge\kern -0.1em V\!/}c^2}\xspace}
\newcommand{\gevgevcccc}{\ensuremath{\gev^2\!/c^4}\xspace} % for q^2

\def\ps   {\ensuremath{\aunit{ps}}\xspace}
\def\fs   {\aunit{fs}}

\def\gsim{{~\raise.15em\hbox{$>$}\kern-.85em
          \lower.35em\hbox{$\sim$}~}\xspace}
\def\lsim{{~\raise.15em\hbox{$<$}\kern-.85em
          \lower.35em\hbox{$\sim$}~}\xspace}

\def\sPlot{\mbox{\em sPlot}\xspace}

\def\mrad{\aunit{mrad}\xspace}
\def\rad{\aunit{rad}\xspace}

\def\tell1  {TELL1\xspace}
\def\ukl1   {UKL1\xspace}

\newcommand{\lhcborcid}[1]{\href{https://orcid.org/#1}{\hspace*{0.1em}\raisebox{-0.45ex}{\includegraphics[width=1em]{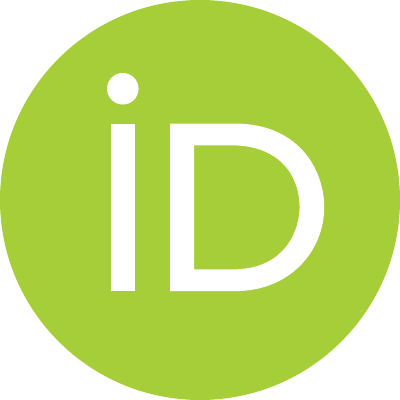}}}}

\hypersetup{
  colorlinks   = true, %Colours links instead of ugly boxes
  urlcolor     = blue, %Colour for external hyperlinks
  linkcolor    = blue, %Colour of internal links
  citecolor    = red   %Colour of citations
}

\ifEnableSectionTOCLinks
    \usepackage[explicit]{titlesec} % to change headings
    
    \let\oldcontentsline\contentsline
    \renewcommand

    \titleformat{\section}{\normalfont\Large\bf}{\hyperlink{tocsection.\thesection}{{\thesection} \parbox[t]{\dimexpr\textwidth-1pc}{#1}}}{1pc}{}

    \titleformat{\subsection}{\normalfont\bf}{\hyperlink{tocsubsection.\thesubsection}{{\thesubsection} \parbox[t]{\dimexpr\textwidth-1pc}{#1}}}{1pc}{}

    \titleformat{name=\section,numberless}[display]{}{}{0pt}{\normalfont\Huge\bfseries #1}
\fi

\usepackage{cite} % Allows for ranges in citations
\usepackage{mciteplus}
\usepackage{longtable} % only for template; not usually to be used in PAPERs

\begin{document}

%%%%%%%%%%%%%%%%%%%%%%%%%
%%%%% Title     %%%%%%%%%
%%%%%%%%%%%%%%%%%%%%%%%%%
\renewcommand{\thefootnote}{\fnsymbol{footnote}}
\setcounter{footnote}{1}

% %%%%%%% CHOOSE TITLE PAGE--------
%\onecolumn
% ===============================================================================
% Purpose: LHCb-INT Note title page template
% Author: P. Koppenburg
% Created on: 2015-05-18
% ===============================================================================

%%%%%%%%%%%%%%%%%%%%%%%%%
%%%%%  TITLE PAGE  %%%%%%
%%%%%%%%%%%%%%%%%%%%%%%%%
\begin{titlepage}

% Header ---------------------------------------------------
\vspace*{-1.5cm}

\noindent
\begin{tabular*}{\linewidth}{lc@{\extracolsep{\fill}}r@{\extracolsep{0pt}}}
\ifthenelse{\boolean{pdflatex}}% Logo format choice
{\vspace*{-1.2cm}\mbox{\!\!\!\includegraphics[width=.14\textwidth]{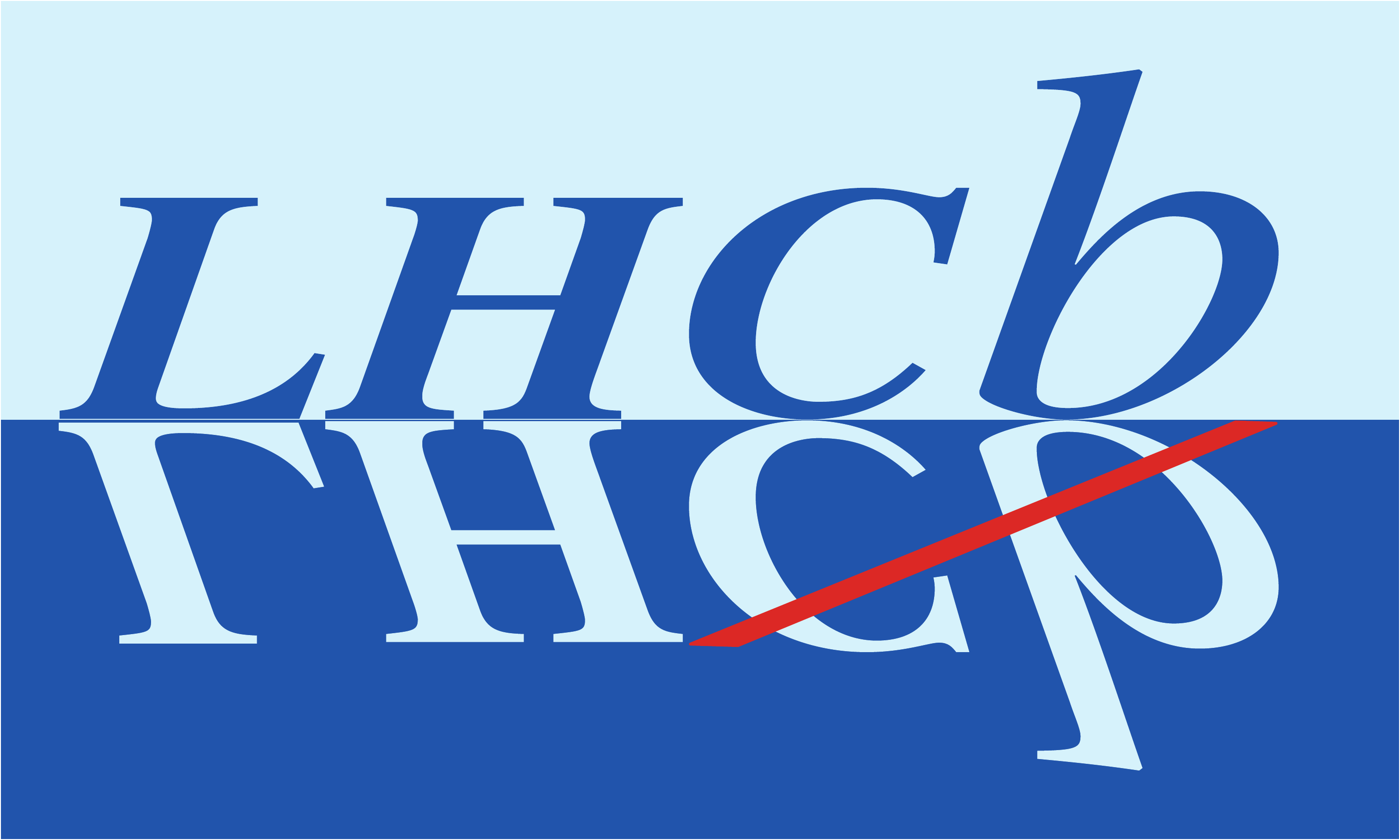}} & &}%
{\vspace*{-1.2cm}\mbox{\!\!\!\includegraphics[width=.12\textwidth]{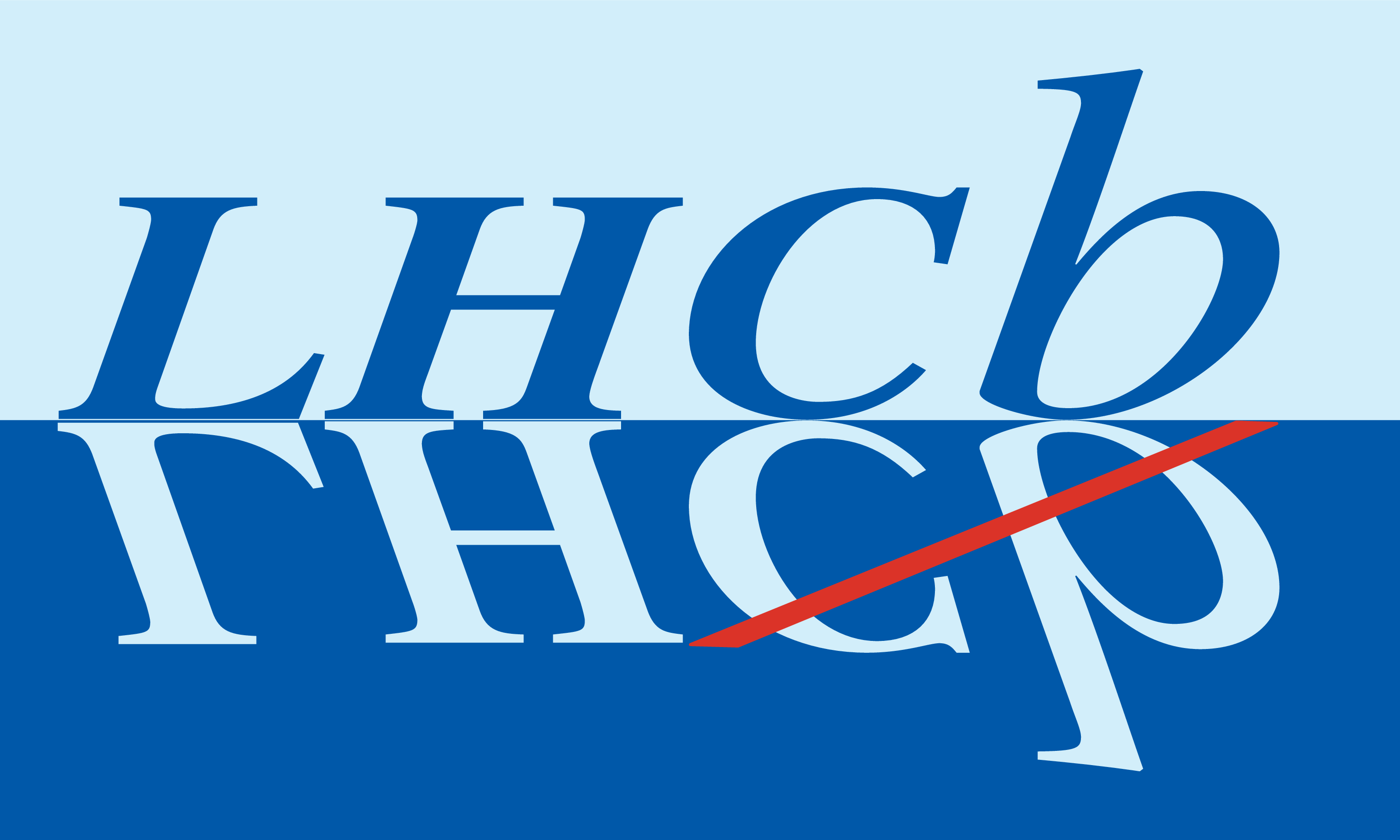}} & &}
 \\
 & & \\  % ID 
 & & 09 October 2026 \\ % Date - Can also hardwire e.g.: 23 March 2010
 & & \\
\hline
\end{tabular*}

\vspace*{4.0cm}

% Title --------------------------------------------------
{\normalfont\bfseries\boldmath\huge
\begin{center}
% DO NOT EDIT HERE. Instead edit macro in main.tex to keep metadata correct
  \papertitle
\end{center}
}

\vspace*{2.0cm}

% Authors -------------------------------------------------
\begin{center}
% If changing to list here, make pdfauthors in main.tex a comma
% separated list with the same names. Otherwise metadata in file will be wrong.
\paperauthors$^1$, on behalf of the \lhcb Collaboration.
\bigskip\\
{\normalfont\itshape\footnotesize
$ ^1$University of Manchester, Manchester, United Kingdom\\
}
\end{center}

\vspace{\fill}

% Abstract -----------------------------------------------
\begin{abstract}
  \noindent 
  These proceedings present the latest measurements of time-dependent properties of \bquark-hadrons using the \lhcb Run 2 data set. Three results are presented. The first of these is a measurement of the CKM phase \phis with \Bs\to\jpsi\pip\pim~\cite{lhcb-paper-2026-017}. This result improves the determination in this channel, obtaining a result of $\phis = -77\pm34(\text{stat})\pm7(\text{syst})$\mrad. This result is included in the \lhcb global average to obtain $\phis = -41\pm17$\mrad. Additionally, a determination of the penguin contributions, $\Delta\phis$, to measurements of \phis using \Bd\to\jpsi\rhoz(770)~\cite{lhcb-paper-2025-059} is presented. This measurement constitutes the first observation of \CP-violation in \Bd\to\jpsi\rhoz decays. The CKM phase $2\beta$ is measured and is determined to be $2\beta^{\text{eff}}_{\cquark\cquarkbar\dquark} = 0.710\pm0.084\pm0.051$\rad. The most stringent constraint on penguin contributions is obtained and determined to be $\Delta\phis = 5.0\pm4.6$\mrad. Finally, the first measurement of time-dependent \CP-violation in a \bquark\to\squark\ellp\ellm process using the decay \Bs\to\KS\mup\mun is presented~\cite{lhcb-paper-2025-062}. The time-dependent coefficients of this decay are determined to be $C=-0.13\pm0.32(\text{stat})\pm0.04(\text{syst})$ and $S=+0.82\pm0.29(\text{stat})\pm0.05(\text{syst})$.
\end{abstract}

\vspace*{2.0cm}
\vspace{\fill}

\end{titlepage}

\pagestyle{empty}  % no page number for the title 

%%%%%%%%%%%%%%%%%%%%%%%%%%%%%%%%
%%%%%  EOD OF TITLE PAGE  %%%%%%
%%%%%%%%%%%%%%%%%%%%%%%%%%%%%%%%

%  empty page follows the title page ----
%\newpage
\setcounter{page}{2}
\mbox{~}

%\input{title-LHCb-ANA}
%\input{title-LHCb-CONF}
%\input{title-LHCb-FIGURE}
%\input{title-LHCb-PAPER}
%\twocolumn
% %%%%%%%%%%%%% ---------

\renewcommand{\thefootnote}{\arabic{footnote}}
\setcounter{footnote}{0}

%%%%%%%%%%%%%%%%%%%%%%%%%%%%%%%%
%%%%%  Table of Content   %%%%%%
%%%%%%%%%%%%%%%%%%%%%%%%%%%%%%%%
%%%% Uncomment if desired
%\tableofcontents

%\cleardoublepage

%%%%%%%%%%%%%%%%%%%%%%%%%
%%%%% Main text %%%%%%%%%
%%%%%%%%%%%%%%%%%%%%%%%%%

\pagestyle{plain} % restore page numbers for the main text
\setcounter{page}{1}
\pagenumbering{arabic}

%% Uncomment during review phase. 
%% Comment before a final submission.#

%% This is the main body
%% It is useful to have a single file so comments are not missed in overleaf.
\section{Introduction}
\label{sec:Introduction}
Presented here are three measurements of the time-dependent (TD) properties of \bquark-hadrons. The measurements were made using the data collected by the \lhcb experiment during its second data-taking period. The \lhcb experiment is a forward-arm spectrometer designed to measure the properties of \bquark- and \cquark-hadrons produced in proton-proton collisions at the Large Hadron Collider. Measurements of the TD properties of \bquark-hadrons are enabled by the \lhcb experiment's high accuracy particle identification, precise vertex resolution, and high granularity tracking. This allows the decay of \bquark-hadrons to be reconstructed with an effective time-resolution better than 50\fs. 

Section \ref{sec:bs2jpsipippim} presents a measurement of the \CP-violating phase \phis using the decay channel \Bs\to\jpsi\pip\pim. Penguin contributions to measurements of \phis must be accounted for. Their effect is constrained using \Bd\to\jpsi\rhoz. The result of this measurement is described in section \ref{sec:bd2jpsirho}. Finally, a measurement of the time-dependent \CP-violation in a \bquark\to\squark\ellp\ellm is presented in section \ref{sec:bdtoksmupmun}. General conclusions and future prospects for TD measurements at \lhcb are presented in section \ref{sec:conclusion}.

\section[CP-Violating Phase phi\_s with B\_s to jpsi pi pi]{
\ensuremath{
    \boldsymbol{C}\!\boldsymbol{P}
    }-violating phase 
\ensuremath{
    \boldsymbol{\phi}_{\boldsymbol{s}}
} with 
\ensuremath{
    \boldsymbol{B}_{\boldsymbol{s}}^{\boldsymbol{0}}
    \boldsymbol{\to} 
    \boldsymbol{J}\mskip -3mu\boldsymbol{/}\mskip -2mu\boldsymbol{\psi}
    \boldsymbol{\pi}^{\boldsymbol{+}} \boldsymbol{\pi}^{\boldsymbol{-}}
}
}
\label{sec:bs2jpsipippim}
The determination of the \CP-violating phase, \phis, is a precision test of the Standard Model (SM). The indirect determination of \phis from fits to the CKM matrix is $-37^{+0.6}_{-0.5}$\mrad \cite{CKMfitter2005}. The \lhcb experiment has made a number of measurements of \phis using a variety of channels. The global \lhcb average from these measurements gives a value of $-31\pm18\mrad$~\cite{lhcb-paper-2023-016}. In order for measurements to reach the precision required to test the SM further measurements are required. 

One of the channels available to measure \phis at \lhcb is the decay \Bs\to\jpsi\pip\pim. Using the Run 1 data set collected in 2011 and 2012, \phis was determined to be $70\pm68\pm8\mrad$ in this decay, where the first uncertainty is statistical and the second is systematic~\cite{lhcb-paper-2014-019}. As the precision of this measurement was statistically limited, further measurements with larger data sets were motivated.

This channel has subsequently been used to measure \phis with the Run 2 data set collected between 2015 and 2018~\cite{lhcb-paper-2026-017}. The measurement begins with the extraction of a sample of signal candidate events from this data set. This is achieved with a combination of selection criteria and a boosted decision tree to reduce the contributions of background processes. The events which pass these requirements are considered to be viable signal candidates. 

To obtain a statistically pure sample, remaining backgrounds are subtracted. The subtraction is performed using the invariant mass of the \jpsi\pip\pim system as an independent variable to obtain signal weights using the \sPlot method. Contributions of known background processes are modelled and fitted to the full spectrum. Each contribution is modelled using a given probability density function. The signal decay and the background \Bd\to\jpsi\pip\pim decays are modelled as a Hypatia distribution plus a Gaussian. Combinatorial backgrounds are modelled using a fifth-order Chebyshev polynomial. The remaining physics backgrounds are modelled using their expected shapes as determined in simulation. The total signal yield determined from this procedure is approximatively 94000.  

In order to perform a TD analysis of this decay, the initial state of the \Bs meson in each event must be determined. The procedure used to achieve this is to utilise a series of flavour tagging (FT) algorithms. These algorithms obtain a tagging decision corresponding to the initial state and a mistag probability, \mistag. FT algorithms are categorised into two types: Opposite Side (OS) and Same Side (SS). OS algorithms determine the \Bds production state by exploiting the partner \bquark from the initial \bquark\bquarkbar pair~\cite{LHCb-PAPER-2011-027,LHCb-PAPER-2015-027}. In contrast to OS algorithms, SS algorithms attempt to tag the signal \bquark using fragmentation tracks from its hadronisation~\cite{LHCb-PAPER-2015-056}. This measurement utilises the OS algorithms and \texttt{SSKaon} algorithm in combination. Before using these algorithms they must be calibrated using a reference channel. The reference channels used for this measurement are \Bp\to\jpsi\Kp and \Bs\to\Dsm\pip for the OS algorithms and \texttt{SSKaon}, respectively. The performance of FT algorithms is measured using their tagging power, \effeff. The power of an algorithm is given by 
\begin{equation}
    \effeff = \etag(1-2\mistag)^2,
\end{equation}
where \etag is the efficiency of assigning a tagging decision to a candidate. In this analysis \effeff after calibration is determined to be $(4.40\pm0.20)\%$, $(4.47\pm0.20)\%$, and $(4.76 \pm 0.22)\%$ in the 2015-2016, 2017, and 2018 samples, respectively.

The \CP-violating phase, \phis, is extracted from the data set using a TD angular fit. In order to perform this fit, effects which arise from the detector and event selection procedure must be accounted for. The first of these effects is the time-resolution of the data set. The time resolution is determined in each event using a calibration from the decay time error. This calibration is determined using a sample of prompt \jpsi\pip\pim decays. The average time resolution is determined to be $40.23 \pm 0.05 \fs$, $37.01 \pm 0.05 \fs$ and $36.93 \pm 0.05 \fs$ in the 2015-2016, 2017, and 2018 samples, respectively. Additionally, the time-dependent selection efficiency is determined using \Bs\to\jpsi\Kstarz decays as a reference channel. Finally, the angular efficiency is determined using simulated \Bs\to\jpsi\pip\pim events.

\begin{figure}
    \centering
    \includegraphics[width=0.55\linewidth]{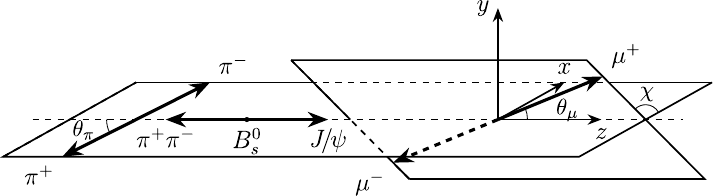}%{Figs/bs2jpsipippim/Fig1.pdf}
    \caption{Illustration of the \Bs\to\jpsi\pip\pim decay planes, which define the decay angles, $\chi$, $\theta_{\pi}$, and $\theta_{\mu}$. Reproduced from \cite{lhcb-paper-2026-017}.}
    \label{fig:decay angles}
\end{figure}

\begin{figure}
    \centering
    \begin{subfigure}{0.37\linewidth}
        \includegraphics[width=\linewidth]{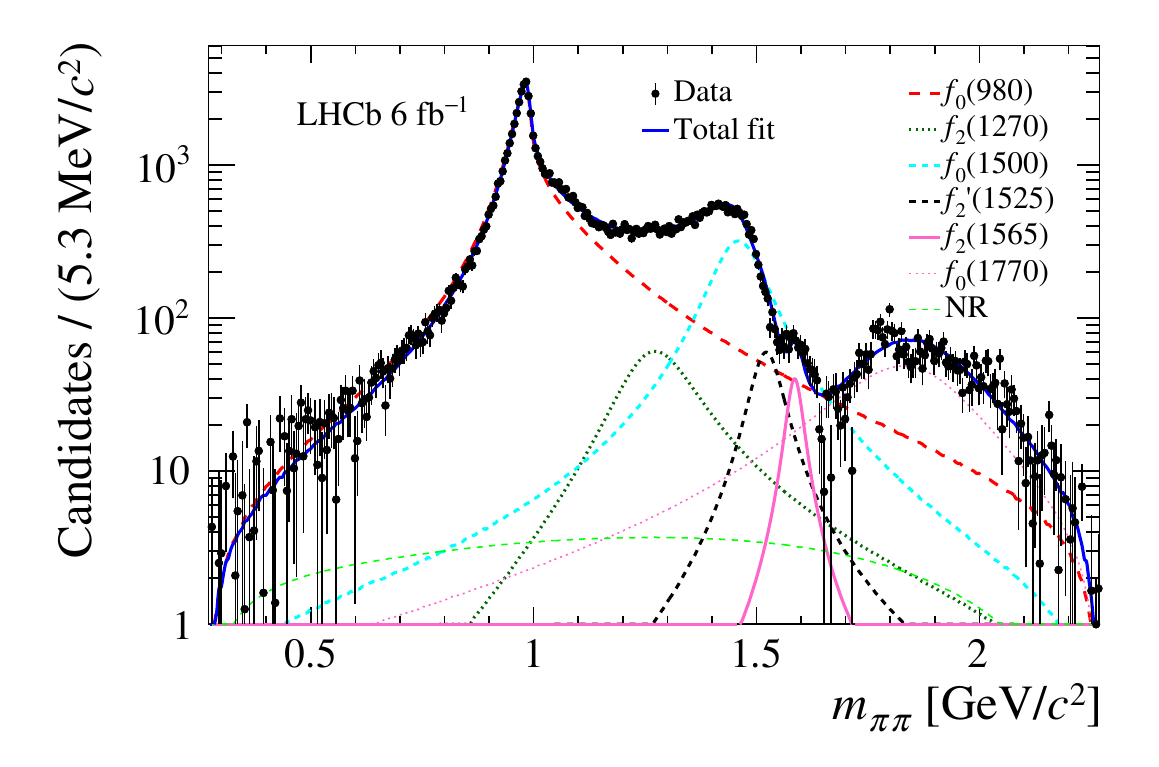}
    \end{subfigure}
    \begin{subfigure}{0.37\linewidth}
        \includegraphics[width=\linewidth]{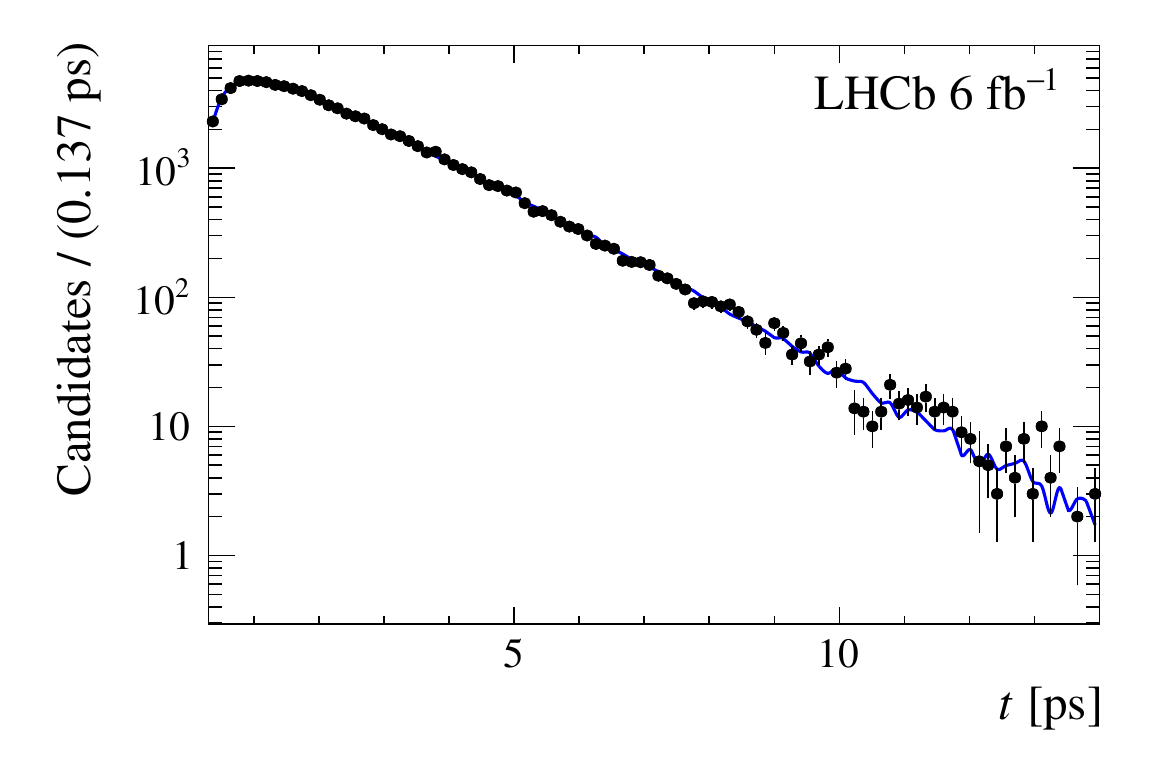}
    \end{subfigure}
    \caption{Distribution of the \pip\pim invariant mass (Left) and the \Bs decay time (Right) in \Bs\to\jpsi\pip\pim decays, with the projection of the fit to these distributions overlaid. Reproduced from \cite{lhcb-paper-2026-017}.}
    \label{fig:bs2jpsipippim angular fits}
\end{figure}

By incorporating the signal weights, flavour tagging, and detector effects, the sample is fitted simultaneously in the decay time, the \pip\pim invariant mass spectrum, and the three decay angles, illustrated in Figure \ref{fig:decay angles}. The projection of the fit result in the \pip\pim invariant mass and \Bs decay time is shown in Figure \ref{fig:bs2jpsipippim angular fits}. The fit obtains $\phis=-0.077\pm0.034\pm0.007$, where the first uncertainty is statistical and the second is systematic. Additionally, the values of the direct CP-violation parameter, $|\lambda|$, and the decay width of the heavy mass eigenstate, $\Gamma_{\text{H}}$, in the \Bs system are measured to be to be $0.610\pm0.002\pm0.004\ps$ and $0.993 \pm0.026\pm0.007$, respectively. The systematic uncertainties on these values are generated by a number of sources, where the systematic uncertainty on the measurement of $\Gamma_{\text{H}}$ is dominated by the decay-time efficiency.

This measurement can be combined with the other \lhcb measurements of \phis to improve the global measurement. The combination of this result and the corresponding measurement with the Run 1 data set gives an average value of $\phis(\Bs\to\jpsi\pip\pim)=-46\pm31\mrad$. Furthermore, the global \lhcb average is improved to $\phis=-41\pm17\mrad$.

\section[Determination of Delta phi\_s with B\_d to jpsi rho]{Determination of \ensuremath{\boldsymbol{\Delta}\boldsymbol{\phi}_{\boldsymbol{s}}} with 
\ensuremath{
    \boldsymbol{B}^{\boldsymbol{0}}
    \boldsymbol{\to} 
    \boldsymbol{J}\mskip -3mu\boldsymbol{/}\mskip -2mu\boldsymbol{\psi}
    \boldsymbol{\rho}^{\boldsymbol{0}}
}(770)}
\label{sec:bd2jpsirho}
Measurements of the \CP-violating phase \phis are susceptible to contamination from penguin processes which can bias the measurement by an amount, $\Delta\phis$. The value of $\Delta\phis$ can be measured using the decays \Bd\to\jpsi\rhoz(770) and \Bs\to\jpsi\Kstarz~\cite{Barel_2021}. The size of $\Delta\phis$ has been estimated using the Run 1 data set to be $0.9\pm9.8$\mrad~\cite{LHCb-PAPER-2014-058}. The precision on measurements of \phis is approaching the scale of the possible penguin pollution, therefore a precise determination of this value is required. 

\begin{figure}
    \centering
    \includegraphics[width=0.6\linewidth]{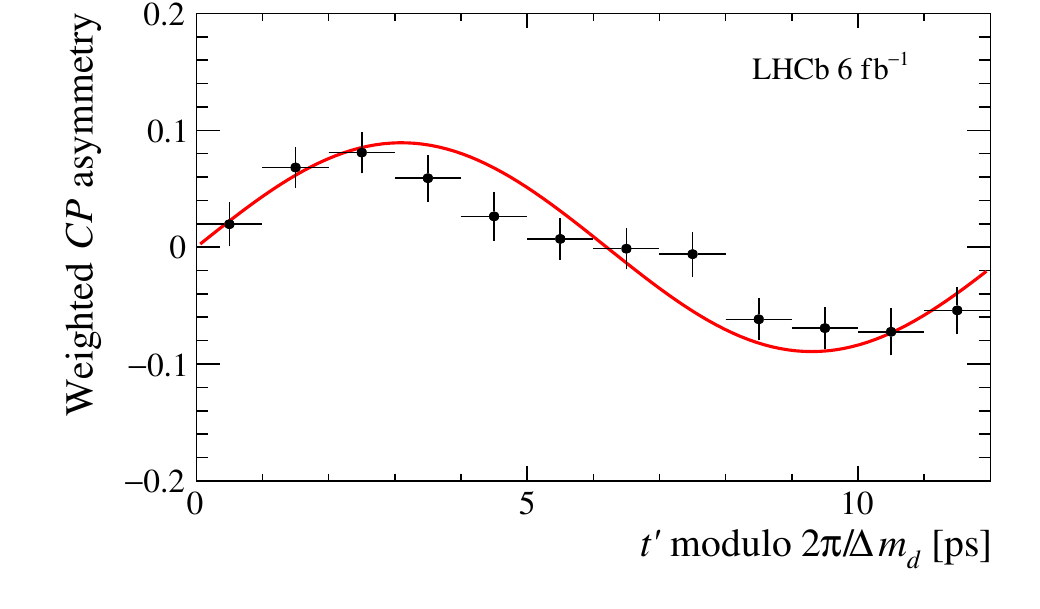}
    \caption{Observed TD \CP asymmetry \Bd\to\jpsi\rhoz(770) decays and the projection of the angular fit to this asymmetry. Reproduced from \cite{lhcb-paper-2025-059}.}
    \label{fig:bd2jpsirho decay time}
\end{figure}

Using the Run 2 data set, the channel \Bd\to\jpsi\rhoz(770) is used to determine the \CP-violating phase $\beta$ and the penguin pollution to \phis~\cite{lhcb-paper-2025-059}. The procedure of this analysis is largely similar to the procedure described in section \ref{sec:bs2jpsipippim}. Notably, as the signal decay contains a \Bd meson, the \texttt{SSPion} and \texttt{SSProton} flavour tagging algorithms are used to tag the \Bd flavour~\cite{LHCb-PAPER-2016-039}. After initial event selection, signal weights are determined using an invariant mass fit. The weights obtained in this fit are used in a fit to the \Bd decay time, \pip\pim invariant mass, and the decay angles. The observed TD \CP asymmetry is illustrated in Figure \ref{fig:bd2jpsirho decay time}. A value of $2\beta_{\cquark\cquarkbar\dquark}^{eff}=0.710 \pm 0.084 \pm 0.051\rad$ is obtained, where the first uncertainty is statistical and the second is systematic. The systematic uncertainty on this measurement is dominated by the exact resonance contributions to the \pip\pim invariant mass spectrum. This measurement constitutes the first observation of \CP-violation in a \Bd\to\jpsi\rhoz decays. 

Using this measurement, a new constraint is placed on the size of the penguin pollution to the measurement of \phis as $\Delta\phis=5.0 \pm 4.6\mrad$. This represents the most stringent constraint of this value. The constraint is determined using an assumption of $SU(3)$ symmetry. If this symmetry is broken, the uncertainty on the value rises to at most $6.4\mrad$. 

\section[Time-Dependent CP-violation in B\_d to KS mu mu]{Time-Dependent \ensuremath{\boldsymbol{C}\!\boldsymbol{P}}-Violation in \ensuremath{
    \boldsymbol{B}^{\boldsymbol{0}}
    \boldsymbol{\to} 
    \boldsymbol{K}^{\boldsymbol{0}}_{\boldsymbol{\mathrm{S}}}
    \boldsymbol{\mu}^{\boldsymbol{+}} \boldsymbol{\mu}^{\boldsymbol{-}}
}}
\label{sec:bdtoksmupmun}
Flavour-changing neutral currents such as \bquark\to\squark\ellp\ellm are suppressed in the SM. As such, these processes are highly sensitive to new physics. Using the full \lhcb Run 1 and 2 data set, this analysis provides a proof-of-principle measurement of the time-dependent \CP violation in a \bquark\to\squark\ellp\ellm process using \Bd\to\KS\mup\mun decays~\cite{lhcb-paper-2025-062}.

\begin{figure}
    \centering
    \includegraphics[width=0.5\linewidth]{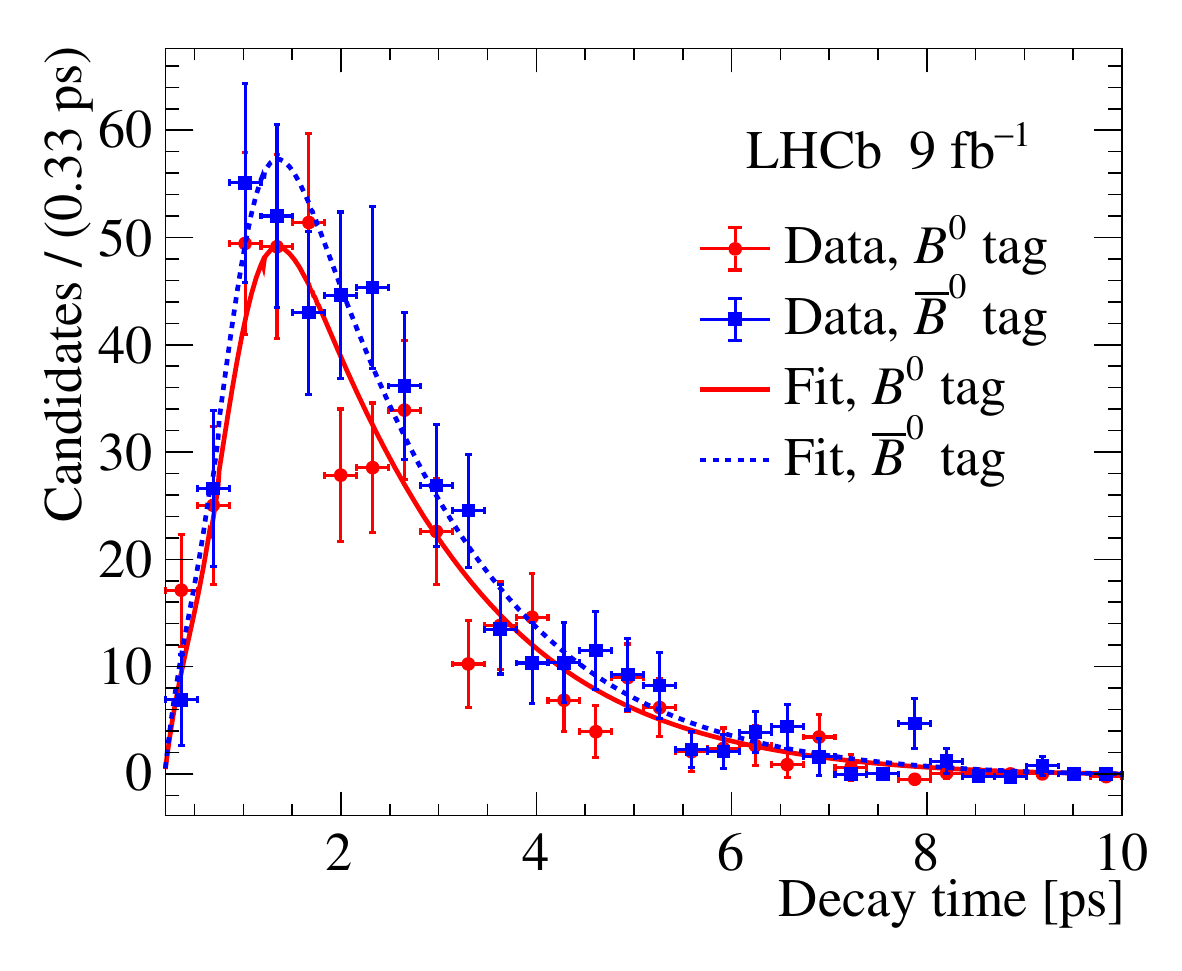}
    \caption{Decay time distribution of \Bd candidates in \Bd\to\KS\mup\mun decays and the corresponding fit to this distribution. Reproduced from \cite{lhcb-paper-2025-062}.}
    \label{fig:bd2ksmupmun decay time}
\end{figure}

\begin{figure}
    \centering
    \includegraphics[width=0.6\linewidth]{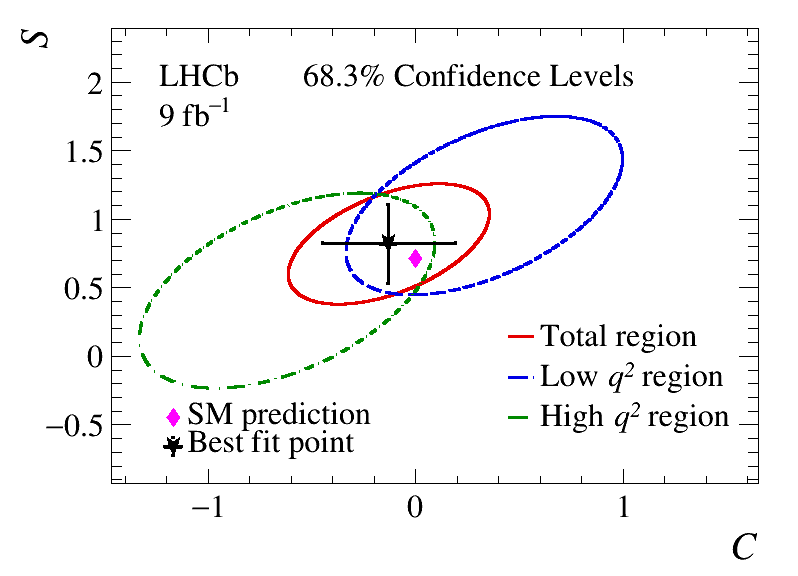}
    \caption{Two-dimensional 68.3\% confidence boundaries of the measurement of $C$ and $S$ in \Bd\to\KS\mup\mun for the full data set and the two $q^2$ bins compared with the SM prediction. Reproduced from \cite{lhcb-paper-2025-062}.}
    \label{fig:bd2ksmupmun CI}
\end{figure}

A sample of signal candidates is obtained by subtracting background contributions using a fit to the \KS\mup\mun invariant mass spectrum. The production flavour of signal candidates is determined using the OS and \texttt{SSKaon} FT algorithms. The measured tagging power in this channel is $2.7\%$ and $4.1\%$ in the Run 1 and Run 2 data sets, respectively. The decay time distribution of the tagged signal candidates is fitted to extract the \CP coefficients $C$ and $S$. From the fit, illustrated in Figure \ref{fig:bd2ksmupmun decay time}, these coefficients are determined to be $C=-0.13\pm0.32\pm0.04$ and $S=+0.82\pm0.29\pm0.05$. Additionally, as potential new physics is expected to feature some dependence on the variable $q^2$, the fit is performed in two bins of this variable. The value of $C$ and $S$ and the definition of these bins are detailed in table \ref{tab:bd2ksmupmun q2bins}. The obtained results of this measurement are consistent with the SM prediction, as shown by the two-dimensional confidence regions in Figure \ref{fig:bd2ksmupmun CI}.

\begin{table}[]
    \centering
    \caption{Values of the \CP coefficients $C$ and $S$ measured in two $q^2$ bins.}
    \begin{tabular}{lcc}
    \hline
    $q^2$ range $[\gevgevcccc]$ & $C$ & $S$ \\
    \hline
    Low: $q^2<8.0$ 
    & $+0.33 \pm 0.44$ & $+1.10 \pm 0.43$\\
    High: $11.0<q^2<15.0$ or $q^2>17.5$
     & $-0.62 \pm 0.47$ & $+0.48 \pm 0.47 $ \\
    \hline
    \end{tabular}
    \label{tab:bd2ksmupmun q2bins}
\end{table}

\section{Prospects and Conclusion}
\label{sec:conclusion}
Presented here are three important measurements for precision tests of the SM. The measurement of \phis in the decay \Bs\to\jpsi\pip\pim obtains a value of $\phis=-77\pm34\pm7\mrad$ and improves the global \lhcb average to $-41\pm17\mrad$. Additionally, using the decay \Bd\to\jpsi\rhoz(770), the effective value of $2\beta$ is determined to be $2\beta_{\cquark\cquarkbar\dquark}^{eff}=0.710 \pm 0.084 \pm 0.051\rad$, representing the first observation of TD \CP violation in this decay. Furthermore, this measurement places the most stringent constraint on the penguin pollution in measurements of \phis, obtaining a value of $5.0 \pm 4.6\mrad$. Finally, the first measurement of TD \CP violation in a \bquark\to\squark\ellp\ellm process was presented using the decay \Bd\to\KS\mup\mun. 

These measurements continue to improve the precision of tests of the Standard Model. Further measurements of these channels are expected in the coming years utilising the \lhcb Run 3 data set which was collected between 2022 and 2026. In addition to this new larger data set, recent development of new FT algorithms which determine a tag from the full underlying event will offer improvements in tagging power of up to 30\%~\cite{LHCb-PAPER-2025-024}. 

\newpage
% Do not include this in any draft (just for information in the template)
%\input{LHCbInternal/acknowledgements_template}
% Comment this in for paper drafts; do not include this in analysis note, conference and figure reports
%\input{acknowledgements}

%\input{LHCbInternal/supplementary_template}

%\input{appendix}

% This should be taken out in the final paper
%\input{supplementary}

% This should be taken out in the final paper
%\input{LHCbInternal/data_and_software_availability_template}

%\addcontentsline{toc}{section}{References}
%\setboolean{inbibliography}{true}
\bibliographystyle{LHCb/LHCb}
\bibliography{main}

%\newpage
%\input{Authorship_LHCb-PAPER-20YY-nnn}
%\input{~/Downloads/Authorship_LHCb-SD-2024-001-grouped-internal-use-only-do-not-publish-2.tex}

\end{document}